# Data Mining: A prediction Technique for the workers in the PR Department of Orissa (Block and Panchayat)

Neelamadhab Padhy[1]  and Rasmita Panigrahi[2]

[1] Assistant. Professor, Gandhi Institute of Engineering and Technology, GIET, Gunupur
`nmp.phdcmj2010@gmail.com, neela.mbamtech@gmail.com`
Research Scholar, Department of Computer Science CMJ University, Meghalaya (Shilong)

[2] Lecturer, Gandhi Institute of Engineering and Technology, GIET, Gunupur
`rasmi.mcamtech@gmail.com`

*Abstract:*

*This paper presents the method of mining the data and which contains the information about the large information about the PR (Panchayat Raj Department) of Orissa .we have focused some of the techniques ,approaches and different methodology's of the demand forecasting .Every organizations are operated in different places of the country. Each place of operation may generate a huge amount of data. In an organization, worker prediction is the difficult task of the manager. It is the complex process not only because its nature of feature prediction but also various approaches methodologies always makes user confused. This paper aims to deal with the problem selection process.In this paper we have used some of the approaches from literatures are been introduced and analyzed to find its suitable organization and situation, Based on this we have designed with automatic selection function to help users make a prejudgment The information about each approach will be showed to users with examples to help understanding. This system also provides calculation function to help users work out a predication result. Generally the new developed system has a more comprehensive functions compared with existing ones .It aims to improve the accuracy of demand forecasting by implementing the forecasting algorithm. While it is still a decision support system with no ability of make the final judgment. .The data warehouse is used in the significant business value by improving the effectiveness of managerial decision-making. In an uncertain and highly competitive business environment, the value of strategic information systems such as these are easily recognized however in today's business environment, efficiency or speed is not the only key for competitiveness. This type of huge amount of data's are available in the form of tera- to peta-bytes which has drastically changed in the areas of science and engineering. To analyze, manage and make a decision of such type of huge amount of data we need techniques called the data mining which will transforming in many fields. We have implemented the algorithms in JAVA technology. This paper provides the prediction algorithm (Linear Regression, result which will helpful in the further research.*

## 1. INTRODUCTION

Data mining is the most important research domain in the 21$^{st}$ century. The objective is to extract the regularities from a huge amount of data .For this purpose some efficient techniques have been proposed like the appriori algorithm.

Forecasting is very important for prediction of the feature events .Science and computer technology together has made significant advances over the past several years and using those advanced technologies and few past patterns, it grows the ability to predict the future .Feature prediction is a process of choosing the best subset of the features available from the selected input data .The best subset contains the minimum number of dimensions that contribute more accuracy.

Accurate demand forecasting remains difficult and challenging in 21$^{st}$ century which is a competitive and dynamic business environment .A little bit improvement in the regard which may gives the significant result to the worker as well manager of an organization. .It aims to improve the accuracy of demand forecasting by implementing the forecasting algorithm. It is a key business function for an organization and employees (workers).the majority of an organization (PR Dept) still conduct demand forecasting using outdated, often homegrown systems that lack forecasting algorithms and analytical tools **[1].**Worker demand in an organization can depend upon various attributes such as size related factors and information about different departments ,and worker attributes.

In our approach, we developed an automated system for attribute classification based on the algorithm. After obtaining the result with help of automated system, we carried out a comparative study of various forecasting technique (Linear Regression) as well as with the help of data mining tool known as WEKA. The data set is used, was collected from the PR department (through the different block head quarters, Orissa) .Rest of this paper focused on the prediction of untested attributes. We have implemented this research with a very sound practical application of Linear Regression technique. The experiments are analyzed with the help of data set to determine their effectiveness when compared with some of the standard stastiscal techniques **[2].**

Data mining, popularly known as Knowledge Discovery in Databases (KDD)**[15]** it is the nontrivial extraction of implicit, previously unknown and potentially useful information from data in databases **[4][5].** Though, data mining and knowledge discovery in databases (or KDD) are frequently treated as synonyms, data mining is actually part of the knowledge discovery process **[3] [5].**

The rapid growth of information technology in various fields of human life has changes in different ways .Information are available in the different formats like records, documents, images, sound recordings, videos, scientific data etc. The data collected from different applications require proper mechanism of extracting knowledge /information from large repositories for better decision making. Knowledge discovery in databases (KDD), often called data mining, aims at the discovery of useful information from large collections of data. The main objective of data mining technique is to extract regularities from a huge amount of data is due to the perception of" we are data rich but information poor". Now a day's workforce scheduling has became and major issue which attract rapid attentions. The reason behind this is modeling the efficient of worker force can not the reduce the cost expenditure of an organization but also it imparts the higher productivity and faster growth(Hunt et al., 2004).The most important step of workforce is prediction of the worker demands for the different positions of the organization. Without accurate estimation of the requirement it is very difficult to generate the standard staff shift patterns, it may happen that we will failed to assign the right tasks to right staff at right time. This may hampers the workforce concept. That's why it is the most important factor in the 21$^{st}$ century which must have to prediction the worker intensive organizations like PR (Panchayat Raj Department, where 341 blocks and 5364 numbers of Gram Panchayat are available) Department, Orissa, Call centers, Retail Industries, NREGA (*National Rural Employment Guarantee Act*), IAY (*Indira Awaas Yojana,* इंदिरा आवास योजना) etc. These are the schemes which provide the PR department of Orissa, not only the Orissa but also in India's most of the states. This workforce scheduling technique which is more suitable in said schemes .In practice, to make accurate estimation of worker demand is rather difficult. Different enterprises have different situations, for some companies, the staffing trends may be relatively stable and it is easy to translate their development strategies into accurate worker demand (Schedule Source, 2009). However, for most companies, the trends change unpredictably, they failed to translate business strategies into the number of worker required to satisfy their objectives (Dan 1996). Furthermore, the continuous competition in the market pushes companies to improve their service quality while minimizing cost expenditures, this add much more pressures to optimize the workforce size (Schedule Source, 2009). As a consequence, worker demand forecasting has been labeled as the weakest process in conducting workforce requirements (Dan 1996).

Demand forecast is one of the most important inputs in the supply chain management .It is basically used manufacturing process ,demands are predicted to perform the basic planning activities such as capacity planning, resource planning etc In difficult and uncertain stages such as demand forecast ,the expert forecast should be integrated with help of the stastiscal methods forecasting[7]

**The objective of this paper is t**o need of assisting the decision making process about companies Workforce size. Therefore, the objectivities separate into two parts. Firstly do research for various worker forecasting approaches in literatures, compare and analysis them to list the strengths and limitations. Analyze examples which applied these approaches to know how they are worked in reality. Try to find the suitable situations of each of the approaches and classify them. After analyzing, try to develop a system to help organization decide their workforce size. The system will have a user interface to show all information of each approach. Customers can directly choose the suitable approaches by reading the descriptions of each approach. Moreover, the system will provide automatically selection and calculation function for users. Users will be required to select an option from a given menu provided on the first tab of the interface. After this selection, the system will generate a series of suggestions which showing at the SUGGESTION tab. The suggestions include a list of the suitable approaches, with each approach a link button.

This paper consists of the six parts, which are introduction, literature review, problem identification and motivation, system design, system implementation and testing, and conclusion. Literature review part will express in detail about the forecasting approaches, their methodology, benefits and limitations. Problem identification part will summaries current conflicts about worker demand forecasting process, analysis the limitations of current existing systems and draw the motivation and requirements of this system. The system design part will describe the design elements in detail with system structure, interface design, data structure design, and algorithm design. System implementation and testing part will display the methods been used to realized the design ideas by analyzing some codes and also show the testing processes. Finally the conclusion part will summaries the achievements; evaluate the system and describe how it could be developed further. The last but not the least which will hugely use for the researchers i.e. Feature scope.

## 2. Literature Review

There are several approaches that have been used for workforce scheduling .In some cases, advances numerical analysis has used for workforce scheduling for workers in an organization like PR Department but in most of the situation clustering technique are used for different types of prediction. There are so many approaches are available but we have used some of the approach.

There has been some few work on this topic since the last two decades. Many researchers have used the forecasting techniques like hotel room arrival forecasting, airline reservation forecasting.

| Topic | et al. | Data Mining forecasting technique |
|---|---|---|
| Data mining Agriculture | [9]Dzeroski,A.Kobler,V.Gjorgijoski,P.Paniov using decision Trees to predict Forest stand Height and Canopy cover from LANDSAT and LIDAR data ,20th Int. conference for environmental Protection-Managing Environmental Knowledge-2006 | WEKA: It is an open source data mining tool. Clustering, Classification, Association Rule. |
|  | [10] "WEKA3: Data mining Software in Java" : Retrieved March 2..7from http://www.cs.waikayo.ac.nz/ml/weka | WEKA tools |
| Soil Data Analysis using classification Techniques and Soil Attribute Prediction | [11] Jay Gholap,Anurag Ingole et al " Soil Data Analysis using classification Techniques and Soil Attribute Prediction" | Classification, Regression, Soil testing in agriculture. Techniques are used : <br> ➢ Naïve Bayes <br> ➢ J48(C4.5) <br> ➢ JRip |
| Multi Relational Data mining Approaches by using demand forecasting | [12]et.al.Dr.Pragnyaban Mishra,Neelamadhab Padhy, Rasmita Panigrahi "MRDM Approaches in the PR. Department of Orissa" published in CIIT Int.Nnational journal of Data mining and Knowledge Engineering ,vol-4 No:55 ,May 2012 | Classification ,clustering ,.Apart from the two data mining models are proposed these are : <br> ➢ Pure Classification model <br> ➢ Hybrid clustering classification model |
| Demand and Forecasting by using fuzzy System | [7]. Armstrong et.al,Green,K.c.,Soo,W " polar Bear population forecast: A public-policy Forecasting Audit",Interfaces,38(2008),3820-405 | Fuzzy clustering ,Neural networks and genetic algorithm |

| Demand forecasting for railway using data mining | [14].Giovanni Melo Carvalho Viglioni " Methodology for Railway demand forecasting using Data Mining" | CRISP-DM,SEMMA,FAYYAD |
|---|---|---|
| Developing the Age Dependent Face Recognition System | [39]. Hlaing Htake Khaung Tin, Myint Sein" Developing the Age Dependent Face Recognition System" | They have predicted the age by using Eigen face for age prediction techniques |

Table 2.1 Literature survey of Different technique

This section will talk about several of human resource forecasting approaches found in the literatures. Currently. Generally, the selected approaches have a broader usability and the methodologies are understandable without specific background knowledge required. This section will mainly describe the seven selected approaches, the first four are quantitative methods which have algorithms to calculate out the worker demand; the last three are qualitative methods which focus on managerial techniques, and the manager plays a significant role (Bohalander & Snell 2007). When we will apply this technique the organization will gain a lot but before that they should have the overall idea about the workers in the organization, its strengths and weaknesses (Noe et al., 2004). Other factors like what they want the organization to be, its size, further business goals should also be clear. This helps to choose the proper worker demand forecasting approaches and builds an optimal workforce size.

## 3. Data Mining System Types [16]

In data mining the classification is one of the most important tasks. We have so many classification approaches for are available for extracting knowledge from data such as statistical [23], Divide-and-conquer [24] and covering [25] approaches. Several approaches has been derived these are Naiave Bayes [26], C4.5 [27], PART [28], Prism [29] and IREP [16]. The different classification methods are used to categorize the data mining methods. One can classify human beings based on their race or can categorize products in a supermarket based on the consumers shopping choices. In general, classification involves examining the features of new objects and trying to assign it to one of the predefined set of classes [31]

The data mining classification system is classified as follows;

> ➢ **What type of classification is to be mined according to the type of data source mined**

The data may classified and handled such as multimedia time-series, test etc
> ➢ **Classification according to the data model**

The classification is based on the data model ,it may be one of the HDM(Hierarchical ),NDM (Network Data Model ),RDM (Relational Data model),ORDM(object Relational Data Model),transactional database, data ware house
> ➢ **Based on what kind of knowledge find**

This classification based on the data mining functionalities, which may be one of the followings:
> ➢ Characterization

- ➢ Discrimination,
- ➢ Association,
- ➢ Clustering

Among different classification attributes we studied but the most attribute is "**Based on what kind of knowledge find**" because this classification presents a clear picture on the different data mining requirements and techniques.

## 4  DIFFERENT APPROACHES

### 4.1 Direct Managerial Input

In this approach we have used that let the headcount or the total cost will be fixed number. Now days it is called as percentage reduction. This approach is helpful to build an optimum staffing template for specific organization (Schedule Source, 2009). As Dan (1996)  .This will helpful when the company has overstaffed  .The primary goals are the expected cash flow or adjustments to the company's return ratios like rate of return, return on capital employed and discounted cash flow return on investment. By using this approach, the future prediction of the can be done and calculated by total figure / average figure and the divisor and dividend could change to meet different analysis criteria like total future output; total investment returns and so on.

**The advantage of this approach is that**
- It is easy to calculate and can implement with simple algorithms.
- This approach can also be used for new established company to make a roughly prediction.

**The disadvantage of this approach is that**
- It does not analysis the business objectives and has no linkage to the actual workload requirement.

### 4.2 Historical Ratio

 According this method by Marching ton and Wilkinson (2005), the overall demand generally related with other business factors .The statistics are as follows [31]
- The  number of items manufactured,
- The number of clients working in the organization ,
- the total budget  of the year

**The advantage of this approach is that**
It is easily understood and can develop with simple methodologies like Excel or Lotus spreadsheets

**The disadvantage of this approach is that**
When we will take any negative value, the productivity remains unchanged because it is very difficult to adjustment for a rapid change of working condition.

## 4.3 Scenario Analysis

This approach is more suitable for the human resource to do the strategic human resource forecasting.

The HR person will take the series of scenarios and by using this scenario they will predict for couple of years. It is one kind of brainstorming sessions with the line manager and the HR manager who have forward looking ability. Let them to develop their scope of the workforce five years or more in the future and then move backwards to find out the key changing points.[32],[33],[34],[35].In general when people discuss the requirements document they are imagination a system does not yet exist. It is generally "What If" analysis ,these cases and sequence of events that results are known as the Scenarios.

The series of actions to be taken as:-

1. Preparation of background

2. Selection of critical indicators

3. Establishing past behavior of indicators

4. Verification of potential future events

5. Forecasting the indicators

6. Writing of scenarios

## 4.4 Regression model Methods

A regression is a statistical analysis [**3**] assessing the association between two variables. It is used to find the relationship between two variables. **et.al [16] Neelamadhab Padhy ,and Rasmita** defined that is a one kind of predictive model which provides the prediction about the unknown data values by using the known data. There are so many techniques are available like Classification, Regression, Time series analysis, Prediction etc. If a set of random data $(x_1, y_1)$ T, $(x_2, y_2)$ T, $(x_n, y_n)$ T for two numerical variables X and Y, where X is a cause of Y. In this linear regression analysis, the distribution of the random data appears as a straight line in X, Y space when X and Y are perfectly related linearly. This captures a relationship between two variables. This line function can be given as [36], [37].

$$\hat{y} = ax + b$$

Here, the linear regression model is used to extract the texture features from the correlation in the frequency channel pairs. The energy values of two frequency channels of one of the channel pair in the top ten list are taken from the channel energy matrix M and consider these energy values as the random data $(x_1, y_1)T, (x_2, y_2)T, \ldots \ldots (x_n, y_n)T$ for two variables X and Y represent a straight line in X,Y space.

The technique regression is basically used in the case of prediction. In Regression analysis the prediction variables are the continuous variables .So many techniques are available Neural Net work, SVM(Support Vector Machine, Linear Regression etc .In this paper we have used the linear regression analysis.

**Regression Formula**

Regression Equation (y) = a + bx slope 'b', Intercept 'a'

$$a = \frac{(\Sigma y)(\Sigma x^2) - (\Sigma x)(\Sigma xy)}{n(\Sigma x^2) - (\Sigma x)^2}$$

$$b = \frac{n(\Sigma xy) - (\Sigma x)(\Sigma y)}{n(\Sigma x^2) - (\Sigma x)^2}$$

$$r = \frac{n(\Sigma xy) - (\Sigma x)(\Sigma y)}{\sqrt{[n\Sigma x^2 - (\Sigma x)^2][n\Sigma y^2 - (\Sigma y)^2]}}$$

And correlation coefficient is 'r'
Where:
x and y are the variables.
b = the slope of the regression line
a = the intercept
point of the
regression line and
the y axis. N =
Number of values or
elements
X = First Score
Y = Second Score
$\Sigma XY$ = Sum of the product of first and Second Scores
$\Sigma X$ = Sum of First Scores
$\Sigma Y$ = Sum of Second Scores
$\Sigma X^2$ = Sum of square First Scores

Using regression algorithms like Linear Regression, Least Square, Simple Regression different attributes were predicted .Acceding to the results the values of the attributes are found most accurately predicted and it depends on least number of attributes.

In such a situation it may happen that all the attributes are numeric, linear regression is a natural and simple technique to consider for numeric prediction, but it suffers from disadvantage of linearity. If the data exhibits the non linear dependency then we may not get the best results .Specially in that cases we may used median regression technique but which incur the high computational cost which is not feasible solution **[8]**

This approach relies on objective data and formal regression models .This is the simplest version among various regression modules is linear regression model.

In this model we have used the time as the independent variable and headcount is the dependent variable Linear regression refers to a model in which the conditional mean of y given the value of x is an affine function of x (Lawrence & Geurts 2008). Most commonly, the relationship between the dependent variable y and all vectors of repressors x is approximately linear. Linear regression model can use the least squares approach which means the overall solution (the line) minimize the sum of the distance between each observed point to the approximate line

In this methodology the basis steps are:

- To collect the related data
- Analysis the data to work out the regression function

**Major Steps**

1. Collect related data, the data could from previous years, or some related data which can be used in an x y coordinate to generate a liner function.

E.g. The PR and HEALTH Department is planning to establish new schemes for old age peoples to provide a medical facility where with one hundred beds are required. They want to know how many Medical Staffs should be recruited so that the hospital can run smoothly. They observed twelve similar sized hospitals nearby, and the result respectively is that: sickbeds number (**t**) [23, 29, 29, 35, 42, 46, 50, 54, 64, 66, 76, and 78]; medical staff number employed by each hospital (**y**) [69, 95, 102, 118, 126, 125, 138, 178, 156, 184, 176, and 225].

2. Use linear regression analysis to work out the function:

$Y = b + a * X$,
$b = (n\Sigma ty - \Sigma t \Sigma y) / [n\Sigma t2 - (\Sigma t) 2]$
$a = y - bt$

E.g. In this case, they need to forecast the number of medical staff needed for the new hospital by the known number of sickbeds. The regression function is where the variable parameter a and b could be generated by using regression analysis method. As a consequence, the regression function is:

$Y = 30.912 + 2.232 * X$.

3. Work out the amount of Y by given X. (Y is the predication)

E.g. Here X = 100 (100 sickbeds), then Y = 30.912 + 2.232 * 100 = 254.063, that means about 254 medical staffs needed.

Due to the requirement of various data, it may not suitable for new established companies and companies entering new market unless they conducted comprehensive surveys to get enough data to establish the regression model. The main strength of this approach is that it can provide organizational insight and usually generate the most accurate requirements (Dan 1996). It works efficiently when there is a long stable history that can be used to reliably analyze the relationships among variables. However, before build the models, it requires long learning of the techniques and the model builders should get fully understand of the techniques and the organization such as its cost budgets, business goals and so on

## 5. Table Description

These are the following table which is used in this paper.

- **Approach Table**

An approach table which consists information about the different approaches attributes inn this tables are ID, name, introduction, strength, limitation and suitability.

- **Approach Application Table** :

This table has the following attributes these are ID, name, description of methodology and examples.

- **System Description Table**

This table stores the description about this system.

Table Structure of Approach Table

Table 5.2

## 6. Linear Regression Algorithm Implementation : (Forecasting Algorithm Implementation)

This is the most important algorithm implementation part and it calculate the most complex calculation part .We have seen the other algorithm are easiest one because these are quantitative Figure…….gives code implementation of Linear Regression Forecasting Method that is included in the packages of „NEELAMADHAB.GIET.llx" and named as „LinearRegression" in class „LinearRegression".

### Code for the Decision Matrix

```
public static final int approachNumbers = 7;
public static final int decisionMatrix[][]=
{{0,0,0,0,0,0,0},
{1,0,0,1,0,0,1},
{1,1,0,1,0,0,1},
{0,1,0,1,1,1,0},
{0,1,1,1,0,0,0},
{0,1,1,1,0,0,0},
{0,1,0,0,0,0,1},
{0,0,0,0,0,0,0}};
public int suitableMethods = 0;
public int whichMethod[] = {0,0,0,0,0,0,0};
```

```
21    public static void LinearRegression1(double[] x, double[] y, int n,
22            double[] a, double[] dt)
23    {
24        int i;
25        double xx, yy, e, f, q, u, p, umax, umin, s;
26        xx = 0.0;
27        yy = 0.0;
28        for (i = 0; i <= n - 1; i++) {
29            xx = xx + x[i] / n;
30            yy = yy + y[i] / n;
31        }
32        e = 0.0;
33        f = 0.0;
34        for (i = 0; i <= n - 1; i++) {
35            q = x[i] - xx;
36            e = e + q * q;
37            f = f + q * (y[i] - yy);
38        }
39        a[1] = f / e;
40        a[0] = yy - a[1] * xx;
41        q = 0.0;
42        u = 0.0;
43        p = 0.0;
44        umax = 0.0;
45        umin = 1.0e+30;
46        for (i = 0; i <= n - 1; i++) {
47            s = a[1] * x[i] + a[0];
48            q = q + (y[i] - s) * (y[i] - s);
49            p = p + (s - yy) * (s - yy);
50            e = Math.abs(y[i] - s);
51            if (e > umax) umax = e;
52            if (e < umin) umin = e;
53            u = u + e / n;
54        }
55        dt[1] = Math.sqrt(q / n);
56        dt[0] = q;
57        dt[2] = p;
58        dt[3] = umax;
59        dt[4] = umin;
60        dt[5] = u;
61    }
```

Figure 6.1 Code implementation of Linear Regression Forecasting

It is worthy to note that, the linear regression algorithm has three input parameters: double typed X and Y series and an integer typed N. Its output parameter is a double array A with two elements A[0] and A[1] so that the linear regression equation can be expressed as Y = A[1] + A[0]*X.

## 7. Designing of an algorithm for the Class methods

Here we are designing the workflow of the class inside the methods .It imparts the process of selecting the most suitable approaches for the specific users.

For selecting the possible users, the system treated as different types of the users may be the public service (Dchool, Govt, Hospital), Organizations, Retailers, manufacturing company etc. After classification, a matchable (suitable) approaches list and the corresponding application requirements should be found for each category .By analyzing comprehensive materials, a series of matchable approaches could be listed for each category. At this stage, a decision table could be developed with the seven approaches horizontally ranked in columns and the six categories vertically ranked in rows.. In the system, a decision matrix (a two dimension array) is defined with the same idea of decision table. The six categories are set as options on HOME tab with each category corresponds a row of the decision matrix. In each row (option) of the matrix, suitable approaches are been marked.

**Pass 1** : In this stage the user will select the option provided the menu. For this purpose the user will select the combo box in the menu.
**Pass 2**: If the user will choose one of these categories the system searches the corresponding row in the   decision matrix and jump to the Pass-3 stage
**Pass 3** : Represent all the Forcasting approaches one after another and link the button
**Pass 4** : In this stage the user must provide the input data and then use the button "RUN The Algorithm" the system will provide the Forcasting results.
**Pass 5**: In this stage, if the user will select the close option of the "JApplet" in any above Passes then the algorithm will stop working

## 8. Implementation

We have used the JAVA programming language under the ECLIPSE software .The entire process is carried out of the two packages, where the first Package contains the class Linear Regression and the other package contains the other classes along with the Applet.

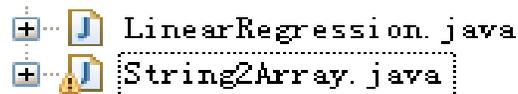

International Journal of Computer Science, Engineering and Information Technology (IJCSEIT), Vol.2, No.5, October 2012

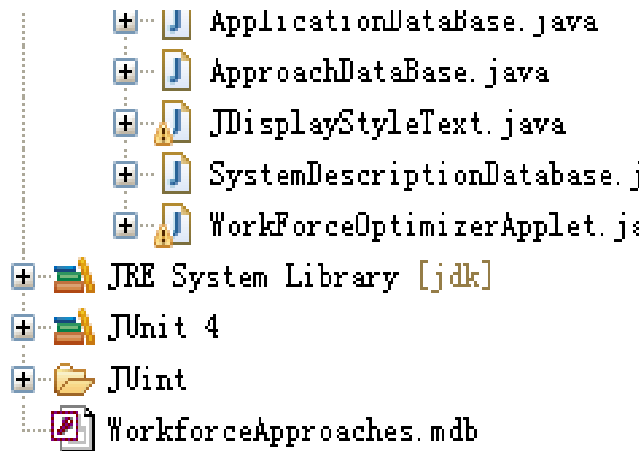

Package Classification and Tree Structure

## 9. Screen Shots

In this paper we have implemented the java programming language ,we have used the java JApplet which contains more number of components like Panel, Button ,Jcombobox etc The main class contains NEELAMADHAB.GIET applet contains huge lines of coding .When we start running the applet then some of the screen shots are displaying .

**C:\Neelamadhab>javac NEELAMADHAB.java**

**C:\Neelamadhab>appletviewer NEELAMADHAB.html**

When we execute the command, then applet will execute and application start running, some of the screen shots are given below.

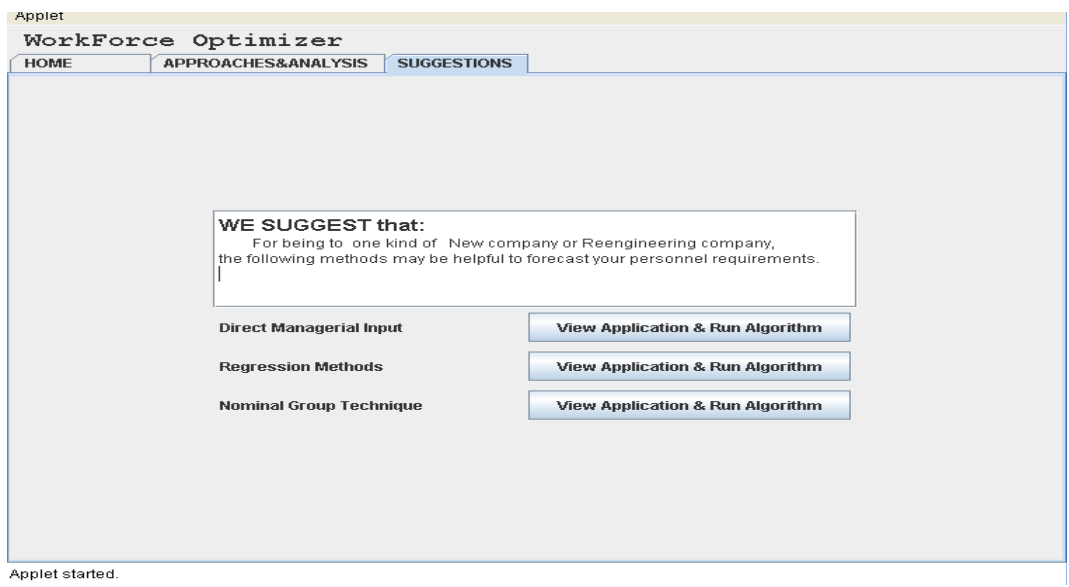

## 10. Comparison of Regression Algorithms

There were very limited variations amongst the predicted values of the approach tables .As the Least Median of square algorithm which produce the better results .We have noticed that the accuracy of the linear regression was s relatively equivalent to that of the least median of squares algorithm

| Algorithm | LRA Linear Regression Algorithm | LMSA (Least Median Square Regression) |
|---|---|---|
| Time taken to build the model | 0.16s | 10.84s |
| Relative Absolute Error | 10.77% | 10.01% |
| Correlation coefficient | 0.9810 | 0.9803 |

Table-2 Describes the Comparison of the result

We have compare the Data mining forecasting technique algorithm and we found that the Relative absolute error is nearly same for both the prediction algorithmns.In the above table it is indicate that LMS regression provides the better result but to build the model is near about 70 times that of Linear regression. So the computational cost used by the LR is lower than that of LMS.

**TESTING OF AN Algorithm FOR CORRECTNESS**

In this section we cross verifies the calculation for the correctness of the different approaches by using the Microsoft Excel or Lotus .Here we are testing whether the results are same or not ?

**Cross verifying the Regression Method**

In this methodology it does the predication by work out a regression function. Here the user will input the series of collected data .

International Journal of Computer Science, Engineering and Information Technology (IJCSEIT), Vol.2, No.5, October 2012

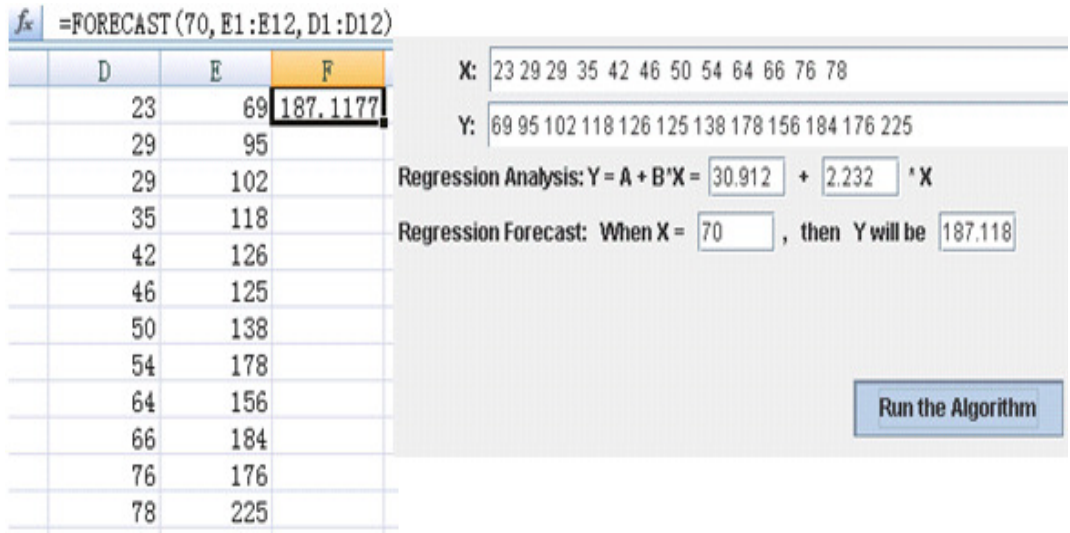

Figure : Regression Method

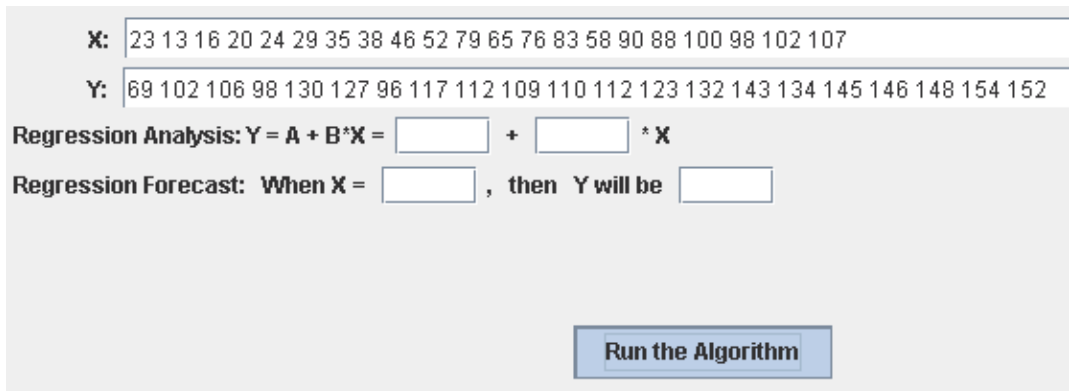

## 11. Conclusion

In this paper we briefly reviewed the various data mining prediction techniques in literature .This research definitely helpful to other researcher to impart the several of data mining prediction issues. It is really very difficult to predict and it is a complex .Actually no approaches or tools can guarantee to generate the accurate prediction in the organization. In this paper, we have analyzed the different algorithm and prediction technique .Inspite the fact that the least median squares regression is known to produce better results than the classifier linear regression techniques from the given set of attributes. As comparison we found that Linear Regression technique which takes the lesser time as compared to Least Median Square Regression.

## 12. Feature Scope

If the researcher will implement the data mining prediction and other classification technique like

K-means,HC(Hybrid clustering ) and PC(Pure classification),Naïve Bayes ,J48(C 4.5),JRip could have been found the better accuracy and minimize the error measure (Root Relative Squared Error).

**Especially For Researchers**

During this implementation of Data Mining Prediction Technique (linear regression) we conclude that linear regression analysis can't handle the large data sets. It doesn't mean that this single model can handle the huge amount of data set . It is highly likely that we need multiple models to fit a large data set.

**To overcome the problems, the researchers can use the two models**

➢ First one -> Parametric Regression Model (Things to remember that the researcher should understand about the complicated large data set )
➢ Second one -> CART (Classification and Regression Tree) decision tree algorithm can be used to build both classification trees (to classify categorical response variables) and regression trees to forecast continuous response variables). Neural network too can create both classification and regression models.

## 13. REFERENCES

[1] Joe Mckendric"Reversing the Supply Chain",Teradata Magazine-Applied Solutions ,Vol.3 Nop.3,2003
[2] L.Armstrong,D.Diepeven & R.Maddern(2004)."The application of data mining techniques to characterize agriculture soil profiles.
[3] De Marco, Jim (2008). "Excel's data import tools". *Pro Excel 2007 VBA*. Apress. p. 43 ISBN 1590599578.
[4] Dunham, M. H., Sridhar S., "Data Mining: Introductory and Advanced Topics", Pearson Education, New Delhi, ISBN: 81-7758-785-4, 1st Edition, 2006
[5] Chapman, P., Clinton, J., Kerber, R., Khabaza, T., Reinartz, T., Shearer, C. and Wirth, R... "CRISP-DM 1.0 : Step-by-step data mining guide, NCR Systems Engineering Copenhagen (USA and Denmark), DaimlerChrysler AG (Germany), SPSS Inc. (USA) and OHRA Verzekeringenen Bank Group B.V (The Netherlands), 2000".
[6] Fayyad, U., Piatetsky-Shapiro, G., and Smyth P., "From Data Mining to Knowledge Discovery in Databases," AI Magazine, American Association for Artificial Intelligence, 1996.
[7] Armstrong et.al,Green,K.c.,Soo,W " polar Bear population forecast: A public-policy Forecasting Audit",Interfaces,38(2008),3820-405
[8] I Written & F.Eibe,(2005),"Data mining practical Machine Learning Tools and technique" 2nd edition ,San Francisco :Morgan Kaufmann
[9] Dzeroski,A.Kobler,V.Gjorgijoski,P.Paniov using decision Trees to predict Forest stand Height and Canopy cover from LANDSAT and LIDAR data ,20th Int. conference for environmental Protection-Managing Environmental Knowledge-2006
[10] "WEKA3: Data mining Software in Java" :Retrieved March 2..7 from http://www.cs.waikayo.ac.nz/ml/weka
[11] Jay Gholap,Anurag Ingole et al " Soil Data Analysis using classification Techniques and Soil Attribute Prediction"

[12] Dr. Pragnyaban Mishra,Neelamadhab Padhy, Rasmita Panigrahi "MRDM Approaches in the PR. Department of Orissa" published in CIIT Int. National Journal of Data mining and Knowledge Engineering ,vol-4 No:55 ,May 2012
[13] Official website of Govt.of India, Ministry of Agriculture http://agricoop.nic.in
[14] Giovanni Melo Carvalho Viglioni " Methodology for Railway demand forecasting using Data Mining"
[15] VSRD-IJCSIT, Vol.2 (3), 2012 pages no -210-222."Data warehousing and its OLAP,MRDM tech. for Decision Support in Business Organization of 21st century"
[16] Neelamadhab Padhy ,Rasmita Panigrahi "Survey of data mining application and Feature scope " , International Journal of Computer Science, Engineering and Information Technology (**IJCSEIT**), Vol.2, No.3, June 2012
[17] Armstrong, M. (2006) *A Handbook of Human Resource Management Practice.* 10st edition. [Internet]
[18] Bajpai, J. N. (1990) *Forecasting the basic inputs to transportation planning at the zonal level.* Transportation Research Board. [Viewed 24th November, 2009].
[19] Chatterjee, S. & Hadi, A. (2006) Regression analysis by example. 4th edition. JOHN WILEY & SONS, INC. Canada. [Viewed 27th March].
[20] Dan, W. (1996) *Workforce Demand Forecasting Techniques.* [Internet] Human Resource Planning, 19 (1): 54-55,
[21] Koontz, H. & Weihrich, H. (2006) Essential of Management. 7th edition. Tata McGraw-Hill Publishing Company Limited. [Viewed 2nd January, 2010].
[22] The Department of Education & Science. (1990) *Projecting the Supply and Demand of Teachers: A Technical Description.* London, HMSO. [Viewed 21st October, 2009].
[23] John, G. H. and Langley, P. (1995). *Estimating Continuous Distributions in Bayesian Classifiers*. Proceedings of the Eleventh Conference on Uncertainty in Artificial Intelligence. pp. 338- 345. Morgan Kaufmann, San Mateo.
[24] Furnkranz, J. (1996). *Separate-and-conquer rule learning. Technical Report* TR-96-25, Austrian Research Institute for Artificial Intelligence, Vienna
[25] Cendrowska, J. (1987). *PRISM: An algorithm for inducing modular rules*. International Journal of Man-Machine Studies. Vol.27, No.4, pp.349-370.
[26] John, G. H. and Langley, P. (1995). *Estimating Continuous Distributions in Bayesian Classifiers*. Proceedings of the Eleventh Conference on Uncertainty in Artificial Intelligence. pp. 338- 345. Morgan Kaufmann, San Mateo.
[27] Quinlan, J. R. (1993). *C4.5: Programs for Machine Learning*. San Mateo, CA: Morgan Kaufmann, San Francisco
[28] Frank, E. and Witten, I. (1998). Generating accurate rule sets without global optimization. In Shavlik, J., ed., *Machine Learning: Proceedings of the Fifteenth International Conference*, pp. 144-151, Madison, Wisconsin, Morgan Kaufmann, San Francisco.
[29] Cendrowska, J. (1987). *PRISM: An algorithm for inducing modular rules*. International Journal of Man-Machine Studies. Vol.27, No.4, pp.349-370.
[30] Furnkranz, J. and Widmer, G. (1994). *Incremental reduced error pruning*. In *Machine Learning: Proceedings of the 11th Annual Conference*, New Brunswick, New Jersey, Morgan Kaufmann.
[31] Grobler, P., Wärnich, S., Carrell, M.R., Elbert, N.F., Hatfield, R.D. (2006) *Human Resource Management in South Africa,* 3rd edition. [Internet].
[32] Krishnamurthi, K. (2006) *Human Resource Management: Concept and Planning.* 1st edition. Global Vision Publishing House. India.
[33] Jacobson, *Object-Oriented Software Engineering: A Use-case Driven Approach*, Addison-Wesley, 1992.
[34] J. Karat and J.L. Bennett, ''Using Scenarios in Design Meetings—A Case Study Example,'' in J.Karat (ed.) *Taking Software Design Seriously: Practical Techniques for Human-Computer Interaction Design*, Academic Press, 1991.

International Journal of Computer Science, Engineering and Information Technology (IJCSEIT), Vol.2, No.5, October 2012

[35] K. Benner, M. Feather, W.L. Johnson and L. Zorman, ''Utilizing Scenarios in the software Development Process,'' *Proc. IFIP WG8.1 Working Conf. Inf. Sys Development Process*, North-Holland, 1993.

[36] M. Lubars, C. Potts and C. Richter, ''Developing Initial OOA Models,''*Proc. 15th Int. Conf.Software Eng.*, IEEE Comp. Soc. Press, 1993.

[37] A.Subha, S.Lenty Stuwart "Linear Regression Model on Multiresolution Analysis for Texture Classification" Published in International Journal of Computer Applications (0975 – 8887) Volume 2 – No.4, June 2010

[38] Shweta Kharya "Using Data Mining Techniques For Diagnosis and Prognosis of Cancer Disease" ,International Journal of Computer Science, Engineering and Information Technology (**IJCSEIT**), Vol.2, No.2, April 2012

[39] Hlaing Htake Khaung Tin, Myint Sein" Developing the Age Dependent Face Recognition System"

## Authors

**Mr. Neelamadhab Padhy** is working as an Assistant Professor in the Department of information and Technology at Gandhi Institute of engineering and Technology (GIET), India. He has done a post- graduate from Berhampur University, Berhampur, India. He is a Life fellow member of Indian Society for Technical Education (ISTE). He is presently pursuing the doctoral degree in the field of Data Mining. He has total teaching experience of 10 years .He has a total of 5 Research papers published in National / International Journals / Conferences into her credit. Presently he has also published 2 Books one is for Programming in C and other is Object Oriented using C++. He has received his MTech (computer science) from Berhampur University Berhampur 2009, His main research interests are Data warehousing and Mining, Distributed Database System. Presently he is pursuing the PhD in computer science under the CMJ University, Shilong, and Meghalaya in the year of 2010.

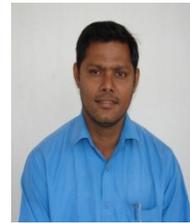

**Mrs.Rasmita Panigrahi** is currently working as a lecturer in the department of information and technology ,Gandhi Institute of Engineering and Technology .She received her MCA from BPUT,(Bijou Patina University of Technology University ,Rourkela 2010 and she has completed MTech(Computer Science) in Berhampur University ,Berhampur . Her main research interests are Data warehousing and Mining, Distributed Database System, Designing and Algorithm. And cryptography .She has published two International/National paper and attended the several conferences.

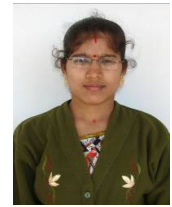